\documentclass[runningheads]{llncs}

\usepackage{subfigure}
\usepackage{graphicx}
\usepackage{calc}
\usepackage{hyperref}

\usepackage{graphics}
\usepackage{footnote}
\usepackage{hyperref}
\usepackage{comment}
\usepackage{xcolor}
\usepackage{amssymb}
\usepackage{amsmath}
\usepackage{tabularx}

\usepackage{float}
\usepackage{booktabs}

\usepackage{ragged2e}
\usepackage{array}

\def\x{\textcolor{blue!70}{\rule{1in}{2in}}}

\def\x{\large DRIVE}
\def\y{\large CHASEDB}
\def\z{\large STARE}

\begin{document}

\title{DR-VNet: Retinal Vessel Segmentation via Dense Residual UNet}

%
%
\author{Ali Karaali\inst{1,2}\orcidID{0000-0002-4154-4181} \and
Rozenn Dahyot\inst{2,3}\orcidID{0000-0003-0983-3052} \and
Donal J. Sexton\inst{1,2,4}\orcidID{0000-0001-8262-632X}}
%
%
\authorrunning{A. Karaali et al.}
%

\institute{The Irish Longitudinal Study on Ageing (TILDA), School of Medicine, Trinity College Dublin, Ireland \and ADAPT: Science Foundation Ireland (SFI) Research Centre for Digital Media Technology, Trinity College Dublin, Ireland  \and Department of Computer Science, Maynooth University, Ireland \and Department of Nephrology, St. James’s Hospital, Dublin Ireland \\
\email{karaalia@tcd.ie}, \email{rozenn.dahyot@mu.ie}, \email{dosexton@tcd.ie}}


\maketitle  

\begin{abstract}
Accurate retinal vessel segmentation is an important task for many computer-aided diagnosis systems. Yet, it is still a challenging problem due to the complex vessel structures of an eye. Numerous vessel segmentation methods have been proposed recently, however more research is needed to deal with poor segmentation of thin and tiny vessels. To address this, we propose a new deep learning pipeline combining the efficiency of residual dense net blocks and, residual squeeze  and excitation blocks. 
We validate experimentally our approach on three datasets and show that our pipeline outperforms current state of the art techniques on the sensitivity metric relevant to assess capture of small vessels. 
\end{abstract}

\section{Introduction}

Retinal fundus images have been widely used as a supportive tool to screen, diagnose  and treat various systemic diseases, such as cardiovascular disorders \cite{Poplin2018}, kidney diseases \cite{kidney1}, and eye-related pathologies, as the retina is the only tissue in the human body where vascular structures can be visualized in a non-invasive manner for  clinical examination. Most of the aforementioned diseases might manifest as changes in the morphological structure of retinal blood vessels, consequently various salient and dangerous diseases (e.g. cardiovascular diseases, high blood pressure) can be detected before they create more dangerous and irreversible conditions. Segmentation of the retinal blood vessels is one of the crucial steps for retinal fundus image analysis. However, segmenting the retinal blood vessels manually is a very time consuming  and  intensive  task,  and requires  not  only  specific  medical training but also technical expertise.  In order to mitigate the workload of the health workers, computerized segmentation strategies  have garnered great interest in recent years, and many segmentation methods have been proposed \cite{CAMARANETO2017182,9054023,channelatt2}. However, one of the main drawbacks of existing retinal vessel segmentation methods is that they present a poor sensitivity rate where the thin and tiny vessel branches are located, and they tend to be miss-classified by most of the existing methods. 

\begin{figure*}[!hb]
\vskip -0.50cm
    \begin{center}
    \subfigure[Overview: Residual Dense-Net (pink blocks) is explained Fig. \ref{fig:densearc},  Residual squeeze and excitation block architecture (blue blocks) is explained Fig. \ref{fig:sqarch}.]
    {
        \includegraphics[width=1.0\textwidth]{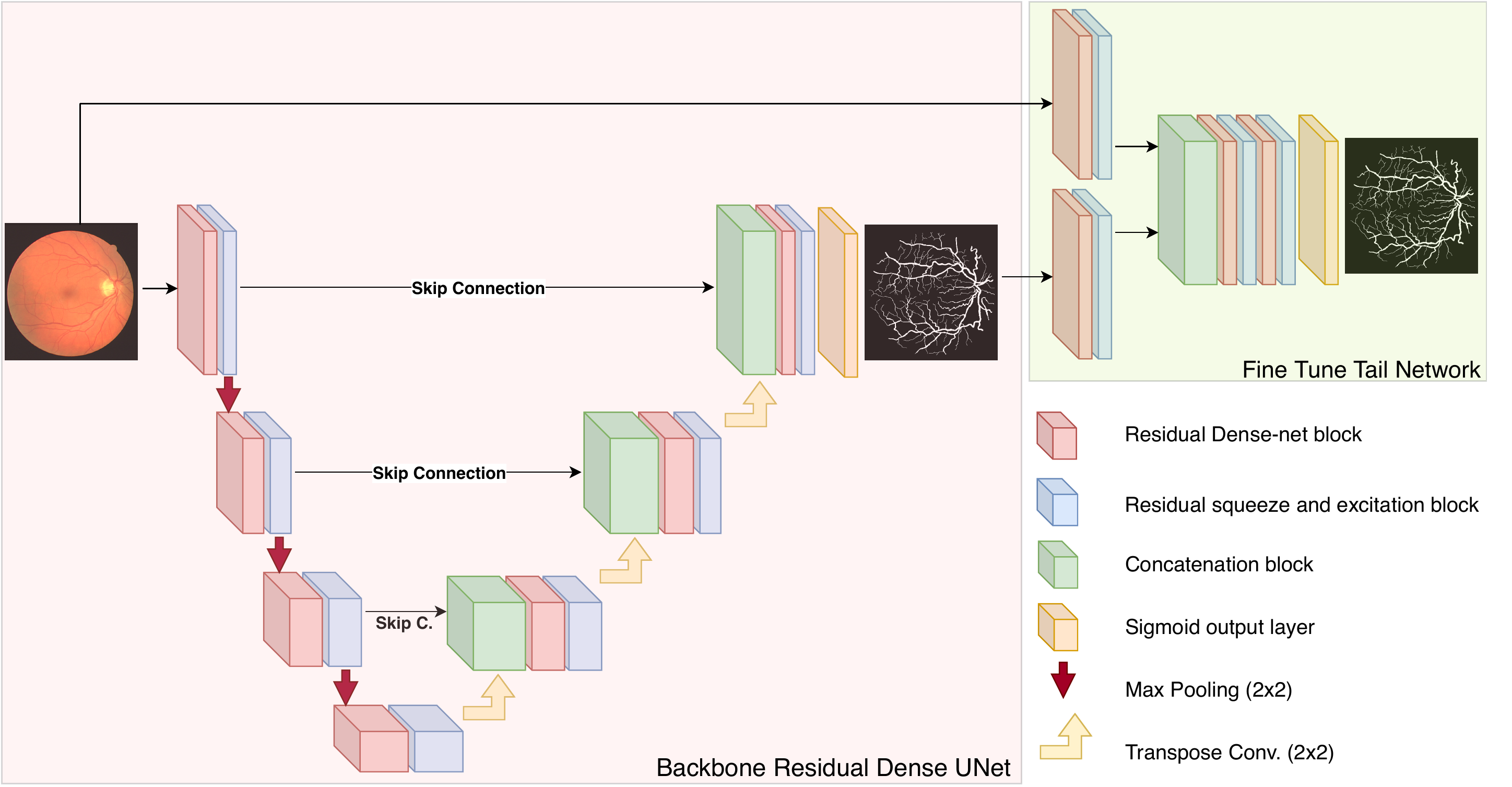}
        \label{fig:archiqutecture}
        }
\vskip -0.20cm
\subfigure[Residual dense-net (RDN) block architecture.]{
     \includegraphics[width=.75\linewidth]{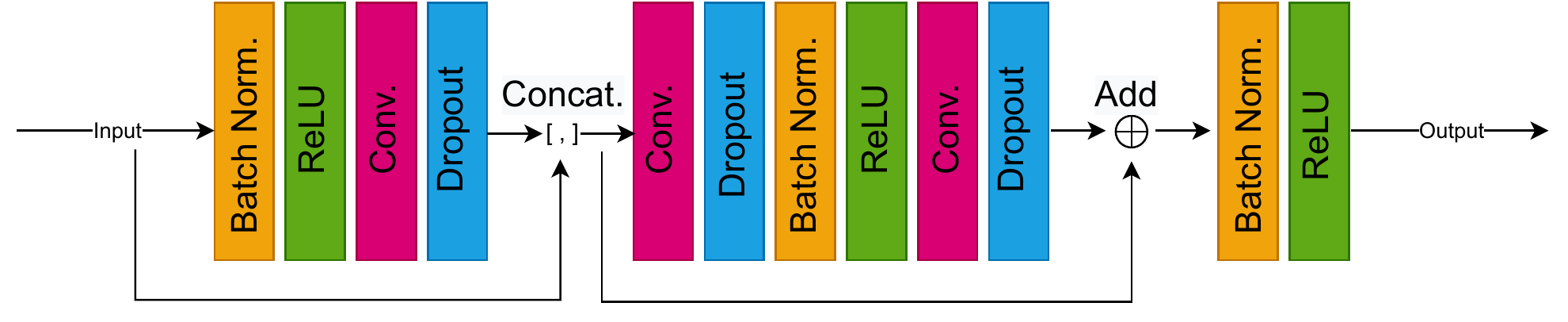}
    \label{fig:densearc} 
}
\vskip -0.250cm
\subfigure[Residual squeeze and excitation block architecture.]{
         \includegraphics[width=.5\linewidth]{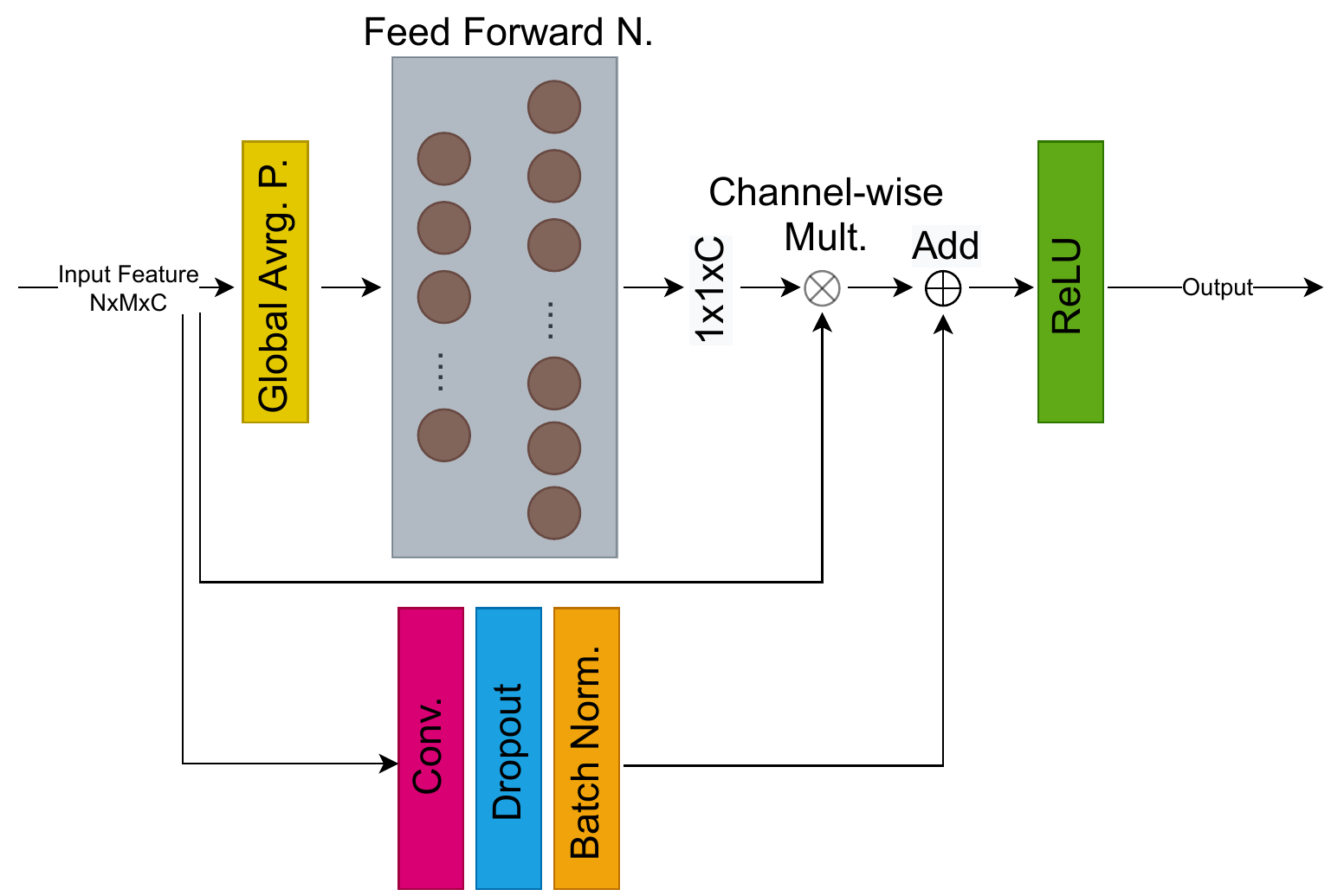}
       \label{fig:sqarch}
}
         \vskip -0.4cm
\caption{Our proposed architecture uses a modified Backbone Residual Dense UNet that is extended with a Fine Tune Tail Network.} \label{fig:overview}
\end{center}
\vskip -0.750cm
\end{figure*}



In  this  paper we  propose a supervised method called DR-VNet for retinal vessel segmentation based on Convolutional Neural Network (CNN), aiming to overcome the aforementioned sensitivity rate problem while keeping the other accuracy metrics (e.g. specificity, accuracy, Area Under the ROC Curve (AUC)) at a high level. The technical details of each part of our pipeline are presented in Section \ref{sec:themethod} and validated on three datasets (cf. Sec. \ref{sec:results}) against the current leading state of the art methods (reviewed  Sec.\ref{sec:soa}).

\section{Related works}
\label{sec:soa}

Existing retinal vessel segmentation methods can be broadly classified into two categories: supervised and  unsupervised methods. 


\textbf{Unsupervised methods} are mostly rule-based methods, and segmentation is carried out by utilizing the visual or geometric information such as contrast levels, vessel structure and other manually designed features \cite{CAMARANETO2017182}.

Bankhead et al. \cite{unsuper2012_01} proposed a method based on a modified wavelet transform. More precisely, their method extract blood vessels by filtering the image with the isotropic undecimated wavelet transform (IUWT) and binarizing the filtered image with a percentile-computed threshold. Nguyen et al. \cite{unsuper_2013_02}'s work, on the other hand, is based on the line operators \cite{4336179}. Their method amplifies blood vessel pixels by filtering the fundus image with a kernel that enhances the pixels that belong to the lines at different orientations. Recently, Li et al.\cite{9054023} proposed a method that is based on a deep tube marked point process (MPP) model \cite{8451108}, which has been originally proposed to detect short and/or long tubes in a given hyper-spectral image. As the blood vessel in retinal fundus images are visually similar to tubes, the authors take advantage of the already trained MPP model bypassing the need for labeled images.

\textbf{Supervised methods} utilize a group of samples to train a classifier that discriminates the vessel pixels from the background eye tissue, which can be referred as a binary classification problem. These  methods are  machine learning and/or deep neural network (DNN) based methods.


Soares et al. \cite{1677727} proposed a supervised segmentation algorithm that utilizes the two dimensional - multi scale Gabor wavelet filter responses as features. Then, three types of classifiers is used to segment vessel pixels: Gaussian mixture model (GMM), K-nearest neighbor(KNN) and least mean square error (LMSE). More recently, Liskowski et al.\cite{deepm01} proposed a CNN based method that consists of consecutive convolutional $\&$ max pooling and fully connected layers. The proposed network inputs $27 \times 27$ image patches, and classifies them whether centered on vessel pixels or not. Moreover, Guo et al. proposed a UNet\cite{unet} based method called Structured Dropout UNet \cite{guo1}. The method is inspired by a recently developed neural block called DropBlock \cite{dropblock}, which does not  exploit the traditional dropout at convolutional layers, but instead,  utilizes a structured dropout block. Their work is extended in \cite{RSAN} by   employing a modified residual block structure and a spatial attention block \cite{spatialatt}. Later on, Guo et al \cite{channelatt2} introduced a new block called Modified Efficient Channel Attention (MECA). Their module enhances the discriminative capacity of the modified UNet shape architecture by weighting the feature map channels independently.


Following the vast majority of recent methods relying on deep learning, we present next a new deep learning pipeline aiming at improving sensitivity for detecting thin blood vessels (see Fig. \ref{fig:overview}).

\section{Proposed method}
\label{sec:themethod}

Our proposed  CNN architecture DR-VNet consists of two cascaded sub-networks: a \textit{Backbone Residual Dense} network and a \textit{Fine-tune Tail} network. 

\textit{The \textit{Backbone Residual Dense}} sub-network is inspired by the well known UNet \cite{unet}, where three down-sampling and up-sampling blocks are used instead of four, and the original convolutional layers are replaced by two novel blocks: (i) Residual dense-net block (Fig. \ref{fig:densearc}); (ii) Residual squeeze and excitation block (Fig. \ref{fig:sqarch}). \textit{The \textit{Fine-tune Tail} sub-network}, on the other hand, is a shallow network that consists of three consecutive RDN and RSE blocks, which combine the output of the \textit{Backbone Residual Dense} sub-network and the input retinal image. The aim is to fine-tune the output of the \textit{Backbone Residual Dense} sub-network to produce the final retinal vessel segmentation image. 


\textbf{Residual dense-net (RDN) block} is a novel neural architecture that is based on DenseNet \cite{densenet} and ResNet \cite{resnet} types of neural structures (see Fig. \ref{fig:densearc}). The proposed neural block consists of two sequentially connected sub-blocks that utilize dense and residual connectivity patterns.

The first sub-block, which employs the dense connectivity pattern, is based on the idea of connecting a layer's output to the following layer's input in a feed-forward fashion,  as in \cite{densenet}. In contrast to summation as in a residual block, the connection is performed by concatenation. To do that, we define a composite function of $\mathcal{H}$ that consists of consecutive Batch Normalization, ReLu, Convolution and Dropout operations, and applied to the input features $X$. More formally, the process can be defined as $y_d = [ \mathcal{H}(X; \{W_i \}), X ]$, where $[.]$ represents the concatenation operation, and $y_d$ is the output of the first sub-block. Although this sub-block might have more than one composite function $\mathcal{H}$ (i.e. a deeper architecture), we opt to use a single unit to keep the network light-weight. 

The second sub-block, on the other hand, employs the residual connectivity pattern, which consists of a mapping function $\mathcal{F}$ with 2-layers of Convolution, Dropout, Batch Normalization, and ReLu operations (see Fig. \ref{fig:densearc}). More precisely, the RDN block is defined as $y_{RDN} = \delta(\mathcal{BN}(\mathcal{F}(y_d; \{W_i \}) + y_d) )$, where $y_{RDN}$ represent the output of the RDN block, $\mathcal{BN}$ represents the Batch Normalization operation, and $\delta$ represents the second non-linearity function ReLu.
This block is designed as a feature extractor inspired by DenseNet and ResNet neural architectures, as they present a good leverage for many CNN tasks \cite{densenet,resnet}. 


\textbf{Residual squeeze and excitation (RSE) block} is inspired by ResNet \cite{resnet} and a recently developed Squeeze and Excitation   (SE) types of neural architectures \cite{sqnetwork}. In general, SE neural blocks aim at exploring channel interdependencies so as to selectively emphasize informative channels and suppress less useful ones. Residual blocks are designed to smooth the information flow across the layers so as to facilitate the optimization process. 
Given the success of  ResNet blocks \cite{resnet} and SE blocks \cite{sqnetwork}, we propose to combine ResNet onto SE block to design a novel neural block (Fig. \ref{fig:sqarch}) to enhance the extracted features and to facilitate the optimization process. 

The proposed RSE block consists of two parallel branches:
\begin{enumerate}
\item  a SE computational branch that transforms the given input $X \in \mathbb{R}^{N' \times M' \times C'}$ to a calibrated feature map $\hat{X} \in \mathbb{R}^{N \times M \times C}$,  
\item and a standard convolutional block with dropout and batch normalization layers that represents the residual mapping function for the RSE block.
\end{enumerate}
More formally, we define the RSE block as $y_{RSE} = \delta( \mathcal{F}(X; \{W_i \}) + \hat{X} )$, where $X$ and $y_{RSE}$ represent the input and output feature maps of the RSE block, $\hat{X}$ represents the output of SE computational branch, $\mathcal{F}(X; \{W_i \})$ illustrates the residual mapping function with weights $W_i$, and $\delta$ is the final non-linearity function ReLu. In this work, we use a single $\mathcal{F}$ composite function of Convolution, Dropout and Batch Normalization respectively.

In the SE computational branch, firstly, the input features undergo a squeeze operation so as to shrink the feature maps across their spatial dimensions and produce a channel descriptor vector $v \in \mathbb{R}^{C'}$ that defines a statistic for each channel of the input features. This is accomplished by using the global average pooling operation:
\begin{equation}
    v_c = \dfrac{1}{N \times M} \sum_{k} \sum_{l} X_c(k,l),
\end{equation}
where $k$ and $l$ represent the spatial locations, $c$ is the channel of interest, $N$ and $M$ are the spatial dimensions. The channel-wise statistic vector $v$ is then re-calibrated through two fully-connected (FC) layers of one hidden layer of size $\tfrac{C}{r}$ with ReLU activations and one output layer of size $C$ with Sigmoid activations. Formally, the recalibration is obtained via a simple gating mechanism $u_c = \sigma(W_2 \ \delta (W_1 v))$, where $\sigma$ and $\delta$ represent the Sigmoid and ReLu activation functions, and $W_1$ $\&$  $W_2$ are the weights of the fully connected layers respectively. Then, the output of the SE branch is obtained by rescaling the input features with the re-calibrated channel-wise statistic vector $\hat{X} = X \otimes u_c$, where $\otimes$ refers to channel-wise multiplication.
This neural block is proposed as a transition block in order to modify the weights of each channel of feature maps so that the informative channels can be emphasized further during the information flow in the network.


\textbf{The network architecture} is designed as a light-weight encoder-decoder type  neural network (Fig. \ref{fig:archiqutecture}). The network contains two consecutive cascaded sub-networks. First sub-network, as noted previously, is based on the famous UNet architecture, which is referred as \textit{Backbone Residual Dense} sub-network; and the second sub-network is based on a shallow CNN, which is referred as \textit{Fine-tune Tail} sub-network. 

The \textit{Backbone Residual Dense} sub-network consists of three down-sampling layers, a latent layer, three up-sampling layers and an output block. Each layer contains a RDN block and a transition RSE block. RDN block is carried out as a feature extractor by using the DenseNet and ResNet connectivity patterns \cite{densenet,resnet}. RSE block is utilized as a transition block that modifies the weights of each channel so as to emphasize the informative ones further. None of these blocks alter the spatial resolution of the input feature maps. Spatial resolution is changed via down-sampling layers by utilizing max-pooling layers of $2 \times 2$ kernel size with 2 pixel stride in the encoder side, and is conducted through transpose convolutions in the decoder side. The output block, which contains a single convolution operation of $1 \times 1$ filters with Sigmoid activation, receives the output of the last up-sampling layer, and yields an initial estimate for the vessel map.

The \textit{Fine-tune Tail} sub-network is a very shallow CNN block, which is designed to fine-tune the initial estimates of the \textit{Backbone Residual Dense} sub-network. The fine-tuning operation is conducted as follows: first, a single-layer of RDN and RSE neural block is applied to the input and output of \textit{Backbone Residual Dense} sub-network; then, outputs are concatenated and sent to a 2-layer of RDN and RSE neural block; and finally, the output block, which has similar architecture as the output block of the backbone sub-network, is fed-forward by the previous layer. The vessel map is then extracted by thresholding.



\section{Experiments}
\label{sec:results}


\textbf{Datasets:} Three publicly available datasets of colour retinal images have been used to evaluate the proposed network.  The DRIVE\footnote{\url{https://drive.grand-challenge.org/}}  dataset \cite{drivedb} consists of 40  retinal  images (resolution 565 $\times$ 584 pixels), which are divided into training and test sets, and each set contains 20 images. 
The CHASE DB\footnote{\url{https://www.kaggle.com/khoongweihao/chasedb1}} dataset \cite{chasedb1} contains 28  retinal images with a resolution of 999 $\times$ 960 pixels, where they are acquired from both the left and right eye of 14 children. The STARE\footnote{\url{https://cecas.clemson.edu/~ahoover/stare/}} dataset \cite{staredb} consists of 20  retinal images with a resolution of 700 $\times$ 605 pixels. Each dataset provides expert annotations for the retinal vessels, and these annotations are used to train the network and to quantitatively evaluate the results.



\noindent \textbf{Implementation details and training procedure:} Given a color retinal fundus image $I$, our algorithm starts with zero padding it in the four margins to a set size $H \times W$. Then, the zero padded images are fed-forward to the retinal vessel segmentation network in order to train the network in two phases:
\begin{enumerate}
    \item In the first phase, we train the \textit{Backbone Residual Dense} sub-network, using an initial learning rate of $10^{-3}$ that is divided by 10 at every 50 epochs, converging after 150 epochs. 
    \item In the subsequent phase, we train the full network by cascading the \textit{Fine-tune Tail} sub-network to the backbone sub-network, using a similar learning rate strategy. In this phase the weights from \textit{Backbone Residual Dense} sub-network are frozen, and the remaining weights in the \textit{Fine-tune Tail} sub-network are learned in 100 epochs.
\end{enumerate}
Our pipeline is implemented using Tensorflow, with a composite loss function that consists of a weighted sum between the binary cross-entropy $L_b$ and the Dice loss $L_d$ \cite{diceloss}:
\begin{equation}
    \mathcal{L} = \lambda_1\ L_b + \lambda_2\  L_d,
    \label{eq:cost}
\end{equation}
where $\lambda_1$ and $\lambda_2$ are the weighting parameters for the loss functions. We used $\lambda_1=1$ and $\lambda_2=0.5$ for both training phases. The scaling ratio for all RSE blocks has been set to $r=2$. As for training time, the convergence is accomplished less than 5 hours on average for each dataset on a Tesla K40m NVidia GPU.

Data augmentation with  random rotations, horizontal, vertical and diagonal flips are applied to all  the training images of the three  datasets. Adam optimizer is utilized with a batch size of $2$ for DRIVE and STARE datasets and a batch size of $1$ for CHASE DB. For the size adjustment, we use $H=W=592$, $H=W=1008$ and $H=W=704$ for DRIVE, CHASE DB and STARE datasets respectively as in \cite{channelatt2}. Finally, the datasets are partitioned as follows: 

\begin{enumerate}
\item  DRIVE, $90$-$10$ \% train-validation data separation is used from the training set of 20 images, and a testing set of 20 images are used for testing; 
\item  CHASE DB, first 20 images are utilized for training-validating ($90$-$10$) \%, and the last 8 images are used for testing; 
\item  STARE, following the previous models \cite{dilatednet,channelatt2}, we adopt a 4-fold cross-validation strategy for training and testing.
\end{enumerate}

The segmented vessel  images (output of our pipeline) are cropped back to the original size, then a thresholding operation is applied, and finally the results are compared with the expert annotations provided with the datasets. 
More precisely, a vessel pixel at location $x$ in the output image $I_o$ is validated if the corresponding pixel location has a higher confidence value than a certain threshold $T$, i.e., if $T < I_o(x)$ (we have set $T = 0.5$). The code and results are shared online\footnote{\url{https://github.com/alikaraali/DR-VNet}}.

\begin{table*}[!t]
\caption{Comparison with state of the art approaches (Vessel-Net \cite{vesselnet}, 2019; DDNet \cite{dilatednet}, 2020; CAR-UNet \cite{channelatt2}, 2021) on the three datasets: DRIVE (\ref{tab:res1}), CHASE DB (\ref{tab:res2}) and STARE (\ref{tab:res3}). Each result is reported as an average over 5 runs for  DRIVE and CHASE DB datasets, and 4-Fold Cross Validation for STARE dataset along with the standard errors. Our pipeline outperforms significantly other approaches systematically for Sensitivity (Se) and G-mean scores, while maintaining excellent performances for Specificity (Sp), Accuracy (Acc) and Area
Under the ROC Curve (AUC).}
\vskip -0.50cm
\scriptsize
\centering
\resizebox{10cm}{!}{
\subtable[\scriptsize{Quantitative evaluations  on the DRIVE dataset.}\label{tab:res1}]{
\begin{tabular}{lccccc}
\hline
\textbf{Method} & \textbf{Sp} $\uparrow$  & \textbf{Se}$\uparrow$  & \textbf{Acc}$\uparrow$  & \textbf{AUC}$\uparrow$  & \textbf{G-mean}$\uparrow$   \\ 
\hline
Vessel-Net \cite{vesselnet}
& 0.9802	& 0.8038	& 0.9578	& 0.9821	& 0.8876 \\
DDNet \cite{dilatednet} 
& 0.9788 &	0.8126 &	0.9594	& 0.9796 &	0.8918 \\
CAR-UNet \cite{channelatt2} 
& \textbf{0.9849} &         0.8135  & \textbf{0.9699} & \textbf{0.9852} &         0.8951  \\ \hline 
DR-VNet 
& 0.9795 & \textbf{0.8512} & 0.9682 & 0.9848 & \textbf{0.9127}     \\ 
stderr \tiny{$\times$ $10^{-3}$}
& 0.5 & 3.2 & 0.2 & 0.5 & 1.4     \\  \hline
\end{tabular}}} \\
\vskip -0.250cm
\resizebox{10cm}{!}{
\subtable[\scriptsize{Quantitative evaluations  on the CHASE DB dataset.}\label{tab:res2}]{
\begin{tabular}{lccccc}
\hline
\textbf{Method} & \textbf{Sp} $\uparrow$  & \textbf{Se}$\uparrow$  & \textbf{Acc}$\uparrow$  & \textbf{AUC}$\uparrow$  & \textbf{G-mean}$\uparrow$   \\ 
\hline
Vessel-Net \cite{vesselnet} 
& 0.9814 & 	0.8186 &	0.9661	& 0.986	& 0.8963 \\
DDNet \cite{dilatednet} 
&0.9773	 & 0.8268 &	0.9637 &	0.9812 &	0.8989 \\
CAR-UNet \cite{channelatt2} 
& \textbf{0.9839} & 0.8439 & \textbf{0.9751} & \textbf{0.9898} & 0.9112  \\  
\hline
DR-VNet
& 0.9733 & \textbf{0.9120} & 0.9694 & 0.9884 & \textbf{0.9421}     \\ 
stderr \tiny{$\times$ $10^{-3}$}
& 0.1 & 6.3 & 1.0 & 0.3 & 2.5     \\ 
\hline
\end{tabular}}} \\
\vskip -0.250cm
\resizebox{10cm}{!}{
\subtable[\scriptsize{Quantitative evaluations  on the STARE dataset.} \label{tab:res3}]{
\begin{tabular}{lccccc}
\hline
\textbf{Method} & \textbf{Sp} $\uparrow$  & \textbf{Se}$\uparrow$  & \textbf{Acc}$\uparrow$  & \textbf{AUC}$\uparrow$  & \textbf{G-mean}$\uparrow$   \\ 
\hline
DDNet \cite{dilatednet} 
& 0.9769& 	0.8391 & 	0.9685 &	0.9858 & 	0.9053 \\
CAR-UNet \cite{channelatt2} 
& \textbf{0.9850} & 0.8445 & \textbf{0.9743} & \textbf{0.9911} & 0.9097  \\  
\hline
DR-VNet 
& 0.9841 & \textbf{0.8572} & 0.9744 & 0.9847 & \textbf{0.9183}     \\ 
stderr \tiny{$\times$ $10^{-2}$}
& 0.1 & 3.4 & 0.2 & 1.0 & 1.8     \\ 
\hline
\end{tabular}}}
\vskip -1.0cm
\end{table*}


\textbf{Quantitative evaluation:} Table \ref{tab:res1}, \ref{tab:res2} and \ref{tab:res3} summarize the average results of a multiple run of our proposed approach along with the standard error and competitive state of the art methods \cite{vesselnet,dilatednet,channelatt2} on the aforementioned three datasets. The following metrics are reported: Specificity (Sp), Sensitivity (Se), Accuracy (Acc), Area Under the ROC Curve (AUC), and the G-mean ($G = \sqrt{Se \times Sp}$) to quantitatively compare our retinal vessel segmentation approach with other competitive methods.

We first illustrate the quantitative results for DRIVE dataset in Table \ref{tab:res1}. As it can be seen from the table, our approach achieves the highest sensitivity rate among the all other approaches ($3.7$ \% better than the second best \cite{channelatt2}), while keeping the other evaluation metrics at a promising level such as specificity, accuracy and AUC. Furthermore, our approach presents the highest G-mean score, which indicates that the proposed approach has a promising trade-off between specificity and sensitivity rates.  
Likewise the quantitative results for CHASE dataset in Table \ref{tab:res2} shows that our proposed approach presents the best performance in terms of Sensitivity rate by presenting $6.8$ \% higher value than the second best \cite{channelatt2} with a significant rise in G-mean score, while keeping the other metrics at a desired level among all competing methods. Notably,  standard errors computed with training our approach  5 times   highlight that our results are significant in showing that our approach provides improvements for these metrics for both datasets.

Finally, the quantitative results for the STARE dataset are summarized in Table \ref{tab:res3}. As it can be seen from the table, the SOTA methods present similar scores on this dataset in terms of used evaluation metrics. However, the proposed approach yields again a significantly higher sensitivity rate and the G-mean score on average when compared to the state of the art methods.

\begin{figure*}[!ht]
    \begin{center}
        ~\x \\
        \includegraphics[width=0.185\columnwidth]{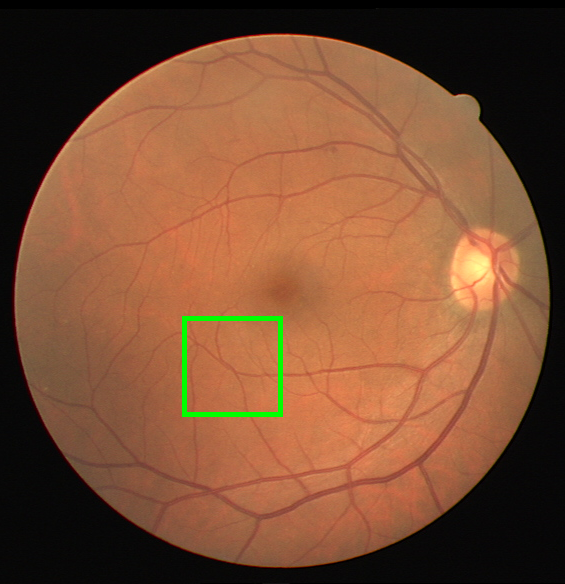}
        \includegraphics[width=0.185\columnwidth]{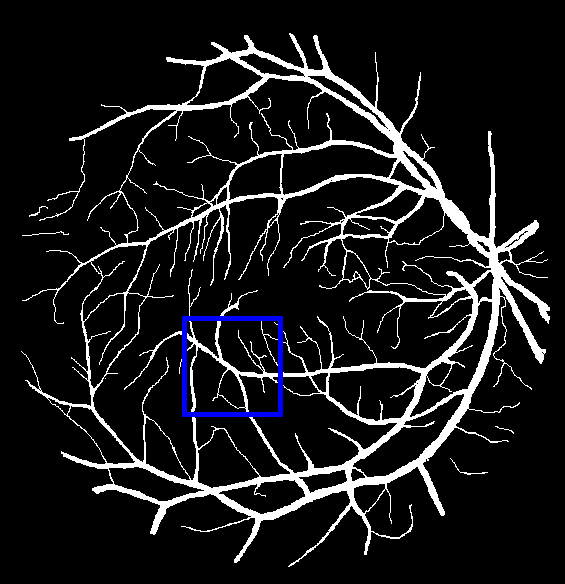}
        \includegraphics[width=0.185\columnwidth]{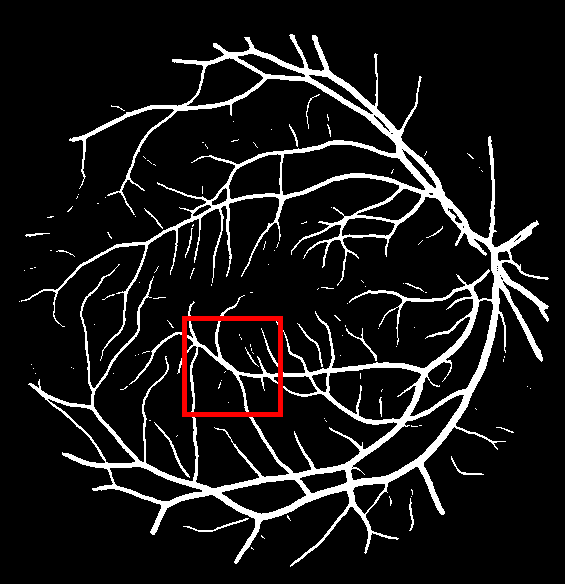} 
        \includegraphics[width=0.185\columnwidth]{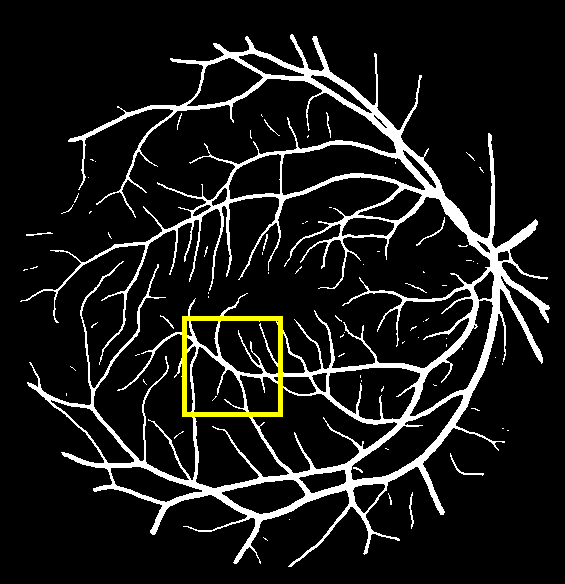} 
        \includegraphics[width=0.185\columnwidth]{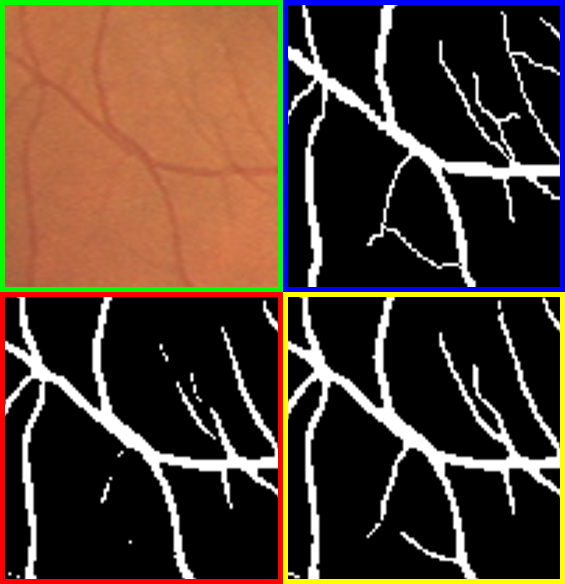} \\
        ~\y \\
        \includegraphics[width=0.185\columnwidth]{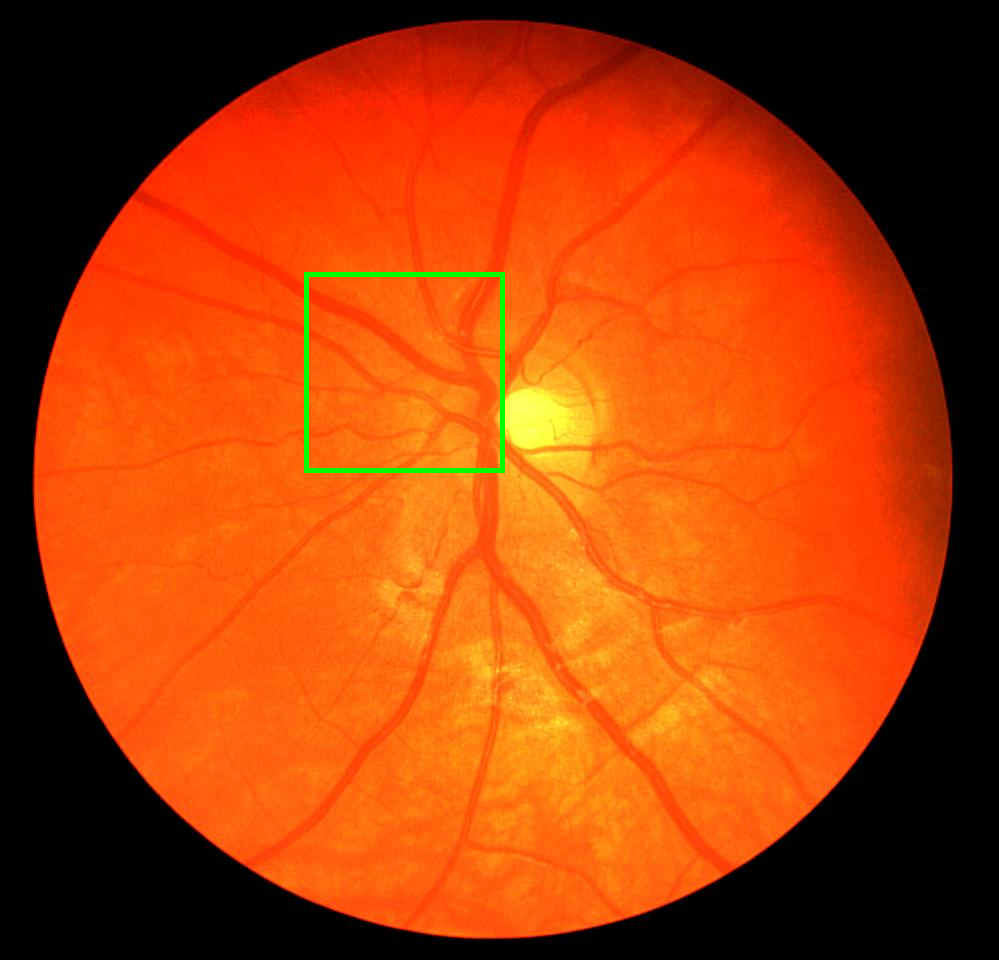}
        \includegraphics[width=0.185\columnwidth]{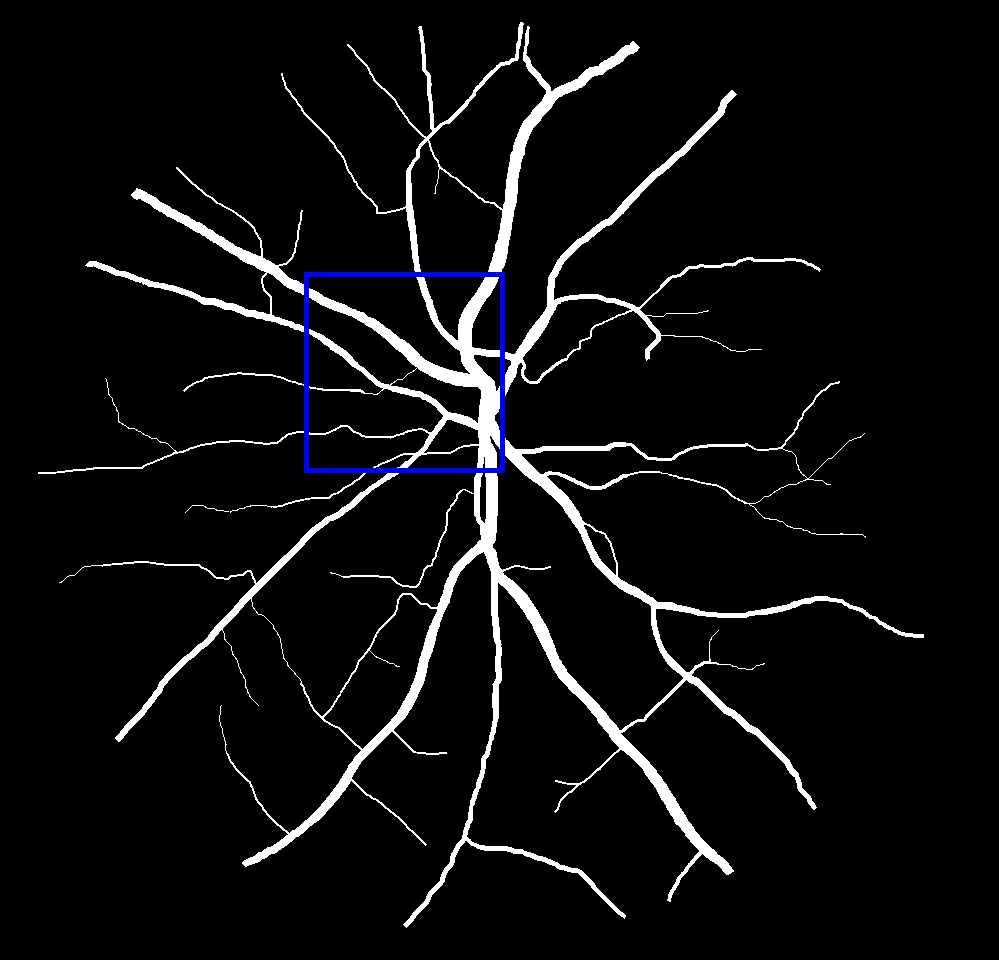}
        \includegraphics[width=0.185\columnwidth]{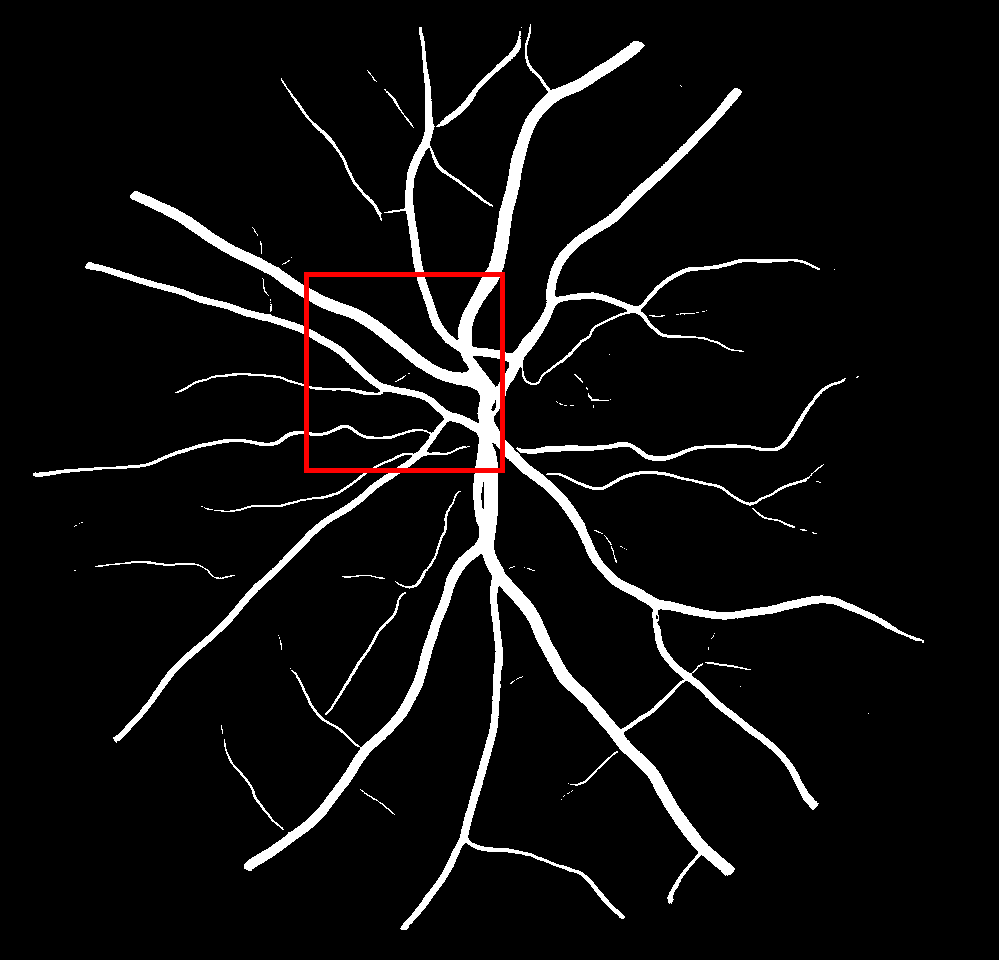}     
        \includegraphics[width=0.185\columnwidth]{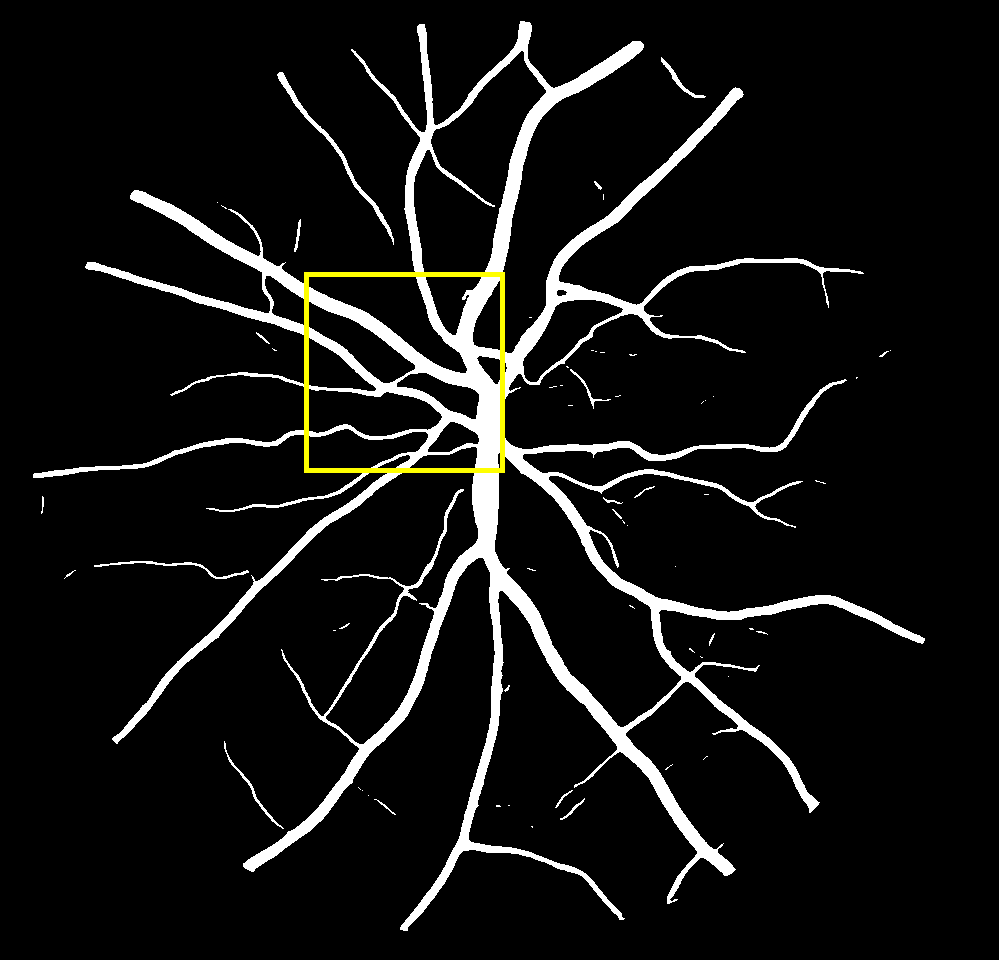}
        \includegraphics[width=0.185\columnwidth]{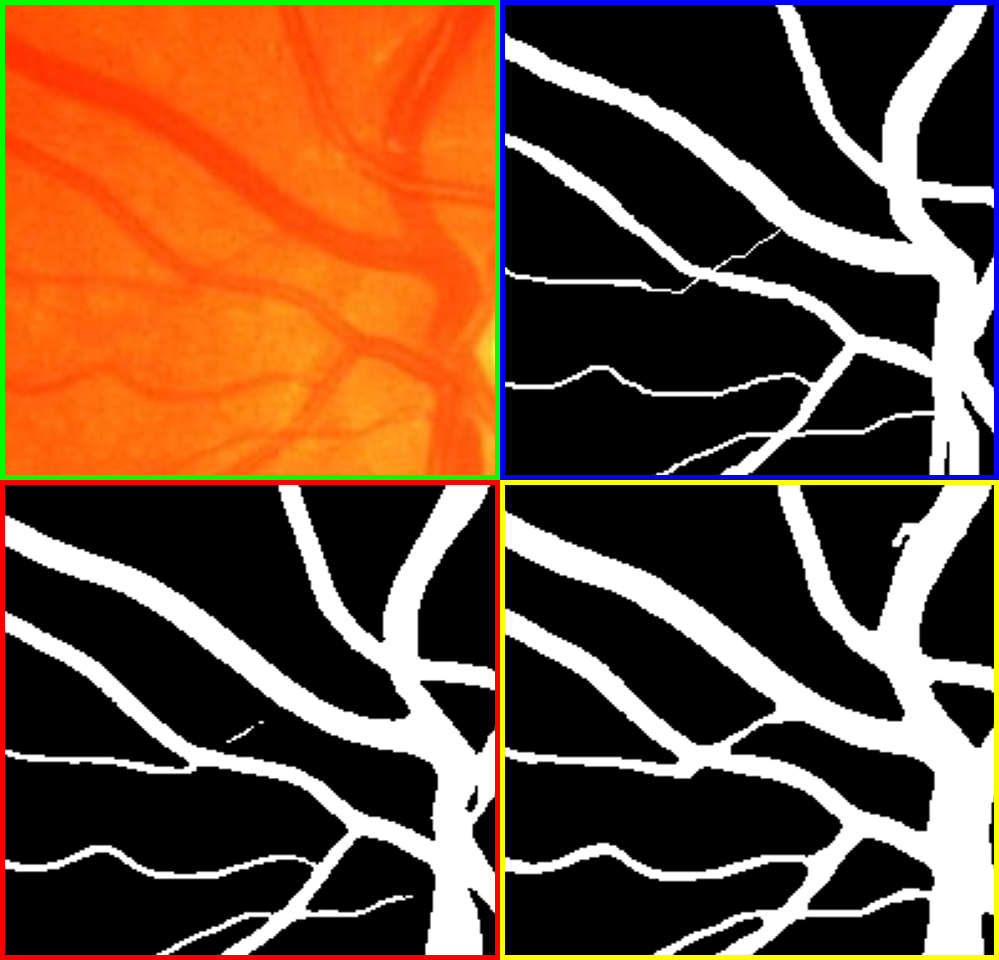} 
        \\
        ~\z \\
        \includegraphics[width=0.185\columnwidth]{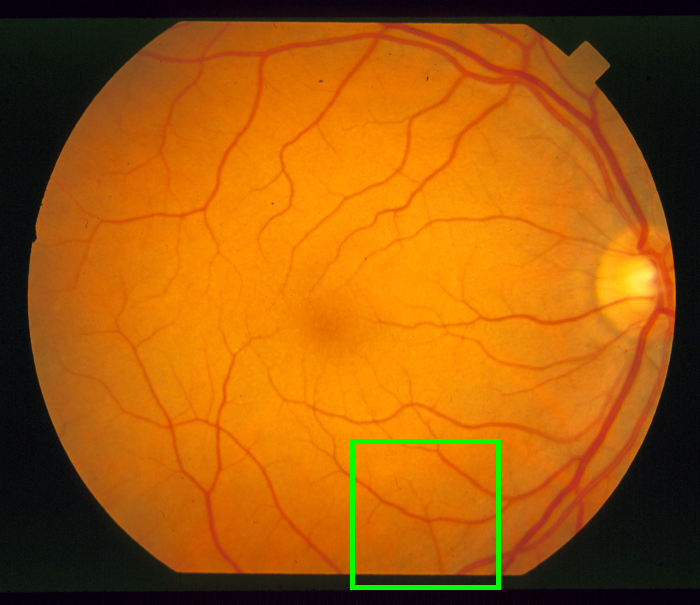}
        \includegraphics[width=0.185\columnwidth]{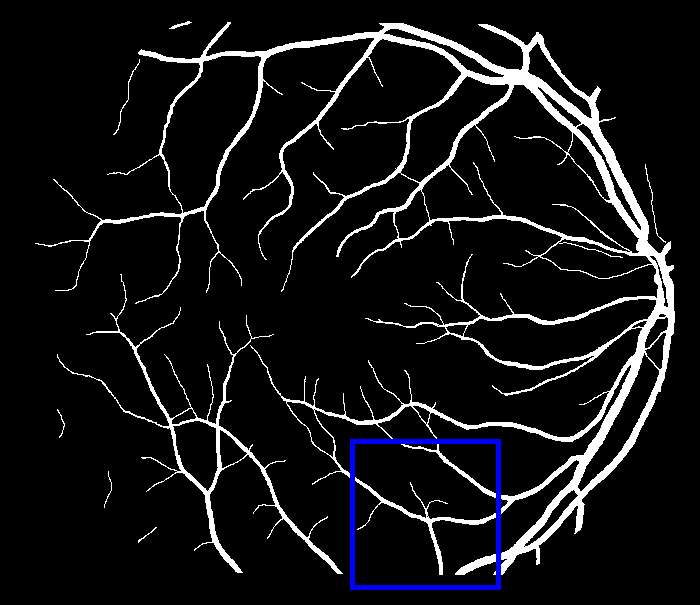}
        \includegraphics[width=0.185\columnwidth]{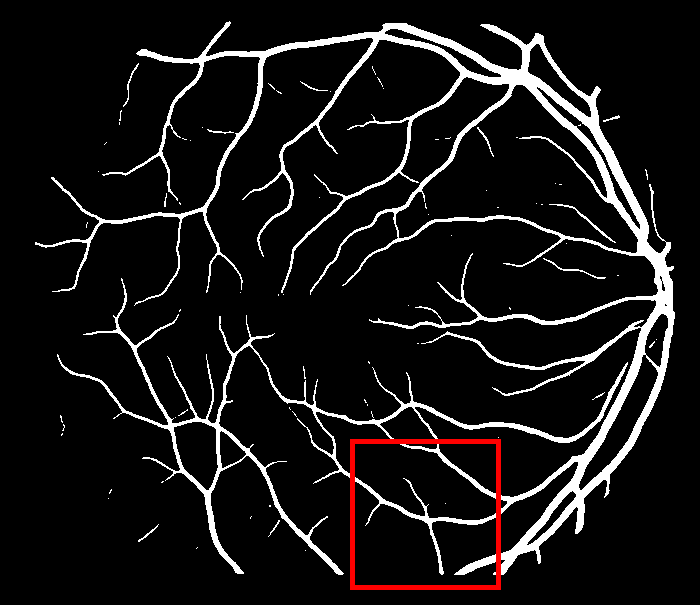}     
        \includegraphics[width=0.185\columnwidth]{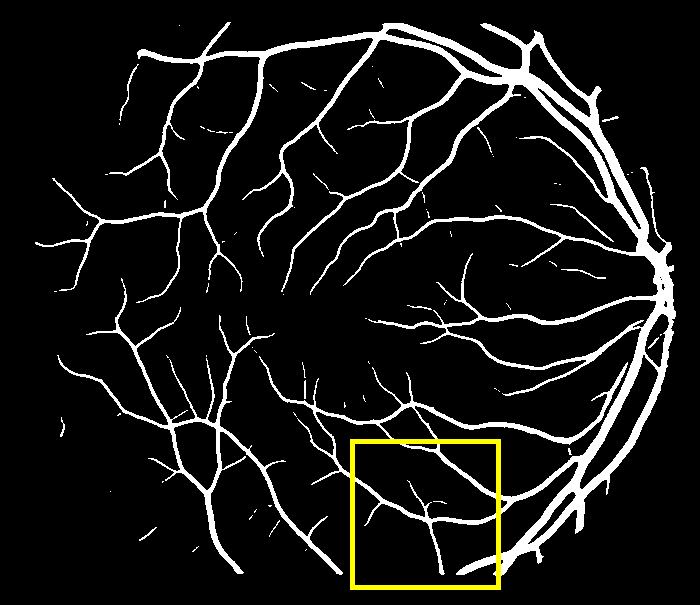} 
        \includegraphics[width=0.185\columnwidth]{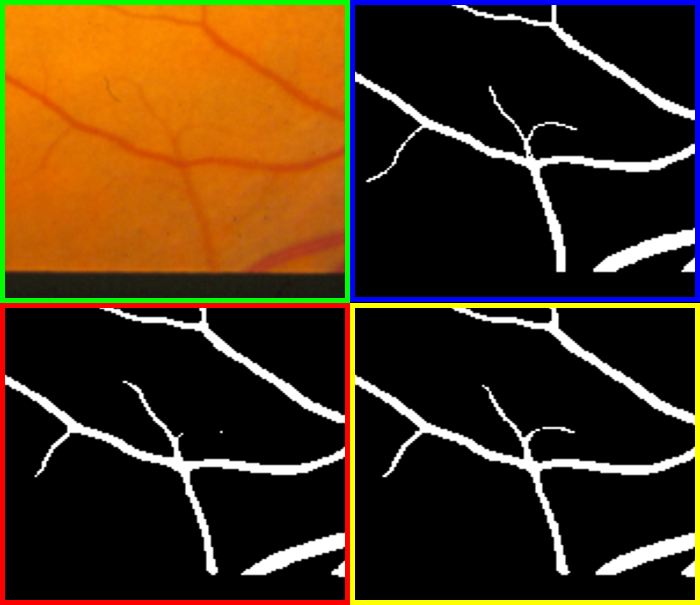} \\
        \begin{tabular}{p{0.2\columnwidth}p{0.185\columnwidth}p{0.2\columnwidth}p{0.185\columnwidth}p{0.185\columnwidth}}
         \small{Retinal image} & \small{Ground truth}& \small{CAR-UNet} \cite{channelatt2} & \small{Ours} & \small{Zoom} \\
        \end{tabular}
             \vskip -0.250cm
        \caption{\label{fig:visualres} Retinal vessel segmentation results on three publicly available datasets. From left to right: Retinal image, ground truth, results of CAR-UNet \cite{channelatt2}, results of the proposed method (Ours), and  zooms of patches for the corresponding images (to read  from left to right as  retinal image (green), ground truth (blue), and in second line,  CAR-UNet \cite{channelatt2} (red) and our proposed method (yellow)).}
    \end{center}
     \vskip -1.0cm
\end{figure*}

The average running times of our approach for a single retinal image at inference are $0.25$s on the DRIVE dataset,  $0.25$s  on the STARE dataset, and  $0.32$s on the CHASE DB dataset. For comparison the average running time for Guo et al's approach~\cite{channelatt2} is $0.35$s on the DRIVE dataset,  $0.35$s  on the STARE dataset, and $0.45$s on the CHASE DB dataset and in~\cite{dilatednet}, the average running time reported is $0.14$s for all datasets using a similar hardware configuration to ours.


\textbf{Qualitative evaluation:} Fig. \ref{fig:visualres} shows the original retinal vessel images, expert annotations for vessels (ground truth), results of the competitive method CAR-UNet  \cite{channelatt2}, and results of the proposed method for the three datasets. Although visual analysis is very subjective, it can be observed from the figure that the proposed approach present visually coherent results by performing a clear segmentation not only of the thick vessels but also of around thin/tiny blood vessels, which can be better seen in the last column of Fig. \ref{fig:visualres}, where the zoomed patches are shown for the original retinal image (upper left), ground truth annotation (upper right), result of the competitive method CAR-UNet \cite{channelatt2} (bottom left), and result of the proposed method (bottom right).

\subsection*{Ablation Study}

The binary cross-entropy is one of the most widely used loss functions, and works very well for many types of binary classification tasks. On the other hand, the Dice loss \cite{diceloss} is a commonly used loss function for medical imaging related segmentation tasks, which has several advantages over binary cross-entropy loss function, such as that the loss information is evaluated both locally and globally. 
We have proposed a composite loss function (cf. Eq. \ref{eq:cost}) that combines the traditional binary cross-entropy loss function with the dice loss and we explore its effectiveness along with the contribution of the fine-tuning sub-network for the DRIVE (largest) dataset. 

We conduct experiments by applying different configuration to the loss function to validate which one presents the best result. First, we train the backbone sub-network using only the binary cross-entropy (BC) loss function;  second, using only the Dice (D) loss function; third, by combining the binary cross-entropy and Dice loss functions (BC\&D); and finally, we trained the backbone and the fine-tuning sub-networks together (i.e. full network) using the proposed composite loss function. It worths noting that experiments are conducted over a single run.

\begin{table}[!t]
\vskip -0.25cm
\caption{Effectiveness of the loss function and fine-tuning sub-block on DRIVE dataset. BC. and D. represent the binary cross-entropy and Dice losses respectively. \label{tab:ablation1}}
\centering
\resizebox{0.8\columnwidth}{!}{
\begin{tabular}{lccccc}
\hline
\textbf{Method} & \textbf{Sp}  $\uparrow$ & \textbf{Se} $\uparrow$ & \textbf{Acc}  $\uparrow$ & \textbf{AUC} $\uparrow$ & \textbf{G-mean}  $\uparrow$ \\ \hline
Backbone with BC
& \textbf{0.9850} & 0.8100          & 0.9696          & 0.9866           & 0.8932  \\  
Backbone with D 
& 0.9802          & 0.8467          & 0.9685          & 0.9471           & 0.9110    \\ 
Backbone with BC\&D 
& 0.9811          & 0.8417          & \textbf{0.9689} & \textbf{0.9868 } & 0.9087     \\ 
Full Net. with BC\&D
& 0.9793          & \textbf{0.8519} & 0.9682          & 0.9855           & \textbf{0.9134}    \\ 
\hline
\end{tabular}
}
\vskip -0.25cm
\end{table}

\begin{table}[!t]
\caption{Effectiveness fine-tuning sub-block on CHASE DB dataset. BC. and D. represent the binary cross-entropy and Dice losses respectively. \label{tab:ablation2}}
\centering
\resizebox{0.8\columnwidth}{!}{
\begin{tabular}{lccccc}
\hline
\textbf{Method} & \textbf{Sp}  $\uparrow$ & \textbf{Se}  $\uparrow$& \textbf{Acc} $\uparrow$ & \textbf{AUC}  $\uparrow$ & \textbf{G-mean}  $\uparrow$ \\ \hline
Backbone with BC\&D. 
& \textbf{0.9815} &	0.8675 &	\textbf{0.9743} &	\textbf{0.9899}	& 0.9227 \\
Full Net. with BC\&D
&  0.9723 & 	\textbf{0.9160} & 	0.9688 & 	0.9887 & 	\textbf{0.9438} \\
\hline
\end{tabular}
}
\vskip -0.250cm
\end{table}

Inference results on the test set are summarized in Table \ref{tab:ablation1}: the Dice loss provides a clear improvement for the sensitivity rate (compare the first and second row in Table \ref{tab:ablation1}), yet there is a trade-off with other metrics, especially with AUC. We also note that there is a significant improvement in performance when the composite loss function is used (see the sensitivity rate at the third row in Table \ref{tab:ablation1}), instead of using either one of them alone. Considering the provided trade-off between sensitivity rate and other evaluation metrics, we opt to use the composite loss function.

Finally, the results obtained by training the full network (freezing the weights of backbone network and training the entire network) are shown at the forth row in Table \ref{tab:ablation1}. It can be observed that there is a slight improvement in sensitivity rate when the fine-tuning sub-network is cascaded to the backbone sub-network. Although the other evaluation metrics remain at a similar level when the network is trained without fine-tuning, the proposed network (i.e. the full network) remarkably yields not only the best balance between the G-mean and AUC scores but also conceivable results with respect to the remaining evaluation metrics. The results of this ablation study are likewise confirmed on the CHASE DB dataset (cf. Tab.
\ref{tab:ablation2}).

\section{Conclusion}

We have proposed a supervised CNN based pipeline for retinal vessel segmentation that is on par with state of the art approaches on three benchmark datasets for metrics such as accuracy, specificity and AUC, but that  outperforms significantly on metrics such G-mean and sensitivity. In practical terms our pipeline is performing better for not only segmenting   the larger and thicker retinal blood vessel branches but also the more distal retinal blood vessel branches which are much smaller and thinner.
Accurate image interpretation for both thick and thin vessels is important since subtle differences in vessel segmentation patterns may be linked with a specific systemic diseases or cardiovascular risk factors. Future work will look at combining information extracted from segmented images with other biological parameters using data from the Irish Longitudinal Study on Ageing (TILDA) \cite{tilda3} in an effort to develop novel prediction tools for cardiovascular disease and other important conditions. 

\noindent \textbf{Acknowledgments:} This work was partly funded by the ADAPT Centre for Digital Content Technology, which is funded under the SFI Research Centres Programme (13/RC/2106\_P2) and is cofunded by the European Regional Development Fund, and also partly supported by Department of Nephrology, St. James’s Hospital, Dublin Ireland. Dr. Donal J. Sexton is funded by Health Research Board of Ireland: grant number ARPP-P-2018-011.

\bibliographystyle{splncs04}
\bibliography{manuscript_v01}

\end{document}